\documentclass[superscriptaddress, twocolumn, appendix, pra, footinbib]{revtex4-2}

\usepackage{graphicx, multirow}
\usepackage{amsmath, amssymb, amsfonts, amsthm, physics, xcolor, hyperref, cleveref, bm, quantikz, xr}
\usepackage{pifont}
\usepackage{colortbl}
\newcommand{\cmark}{{\color{red}YES}}%
\newcommand{\xmark}{{\color{green!60!black}NO}}%
\usepackage{lipsum}
\usepackage{babel, comment, textpos}
\newcommand{\aqa}{$\langle {\rm aQa}\rangle ^L $ Applied Quantum Algorithms, Universiteit Leiden}
\newcommand{\lorentz}{Instituut-Lorentz, Universiteit Leiden, Niels Bohrweg 2, 2333 CA Leiden, Netherlands}
\newcommand{\liacs}{LIACS, Universiteit Leiden, Niels Bohrweg 1, 2333 CA Leiden, Netherlands}
\newcommand{\cern}{CERN, Espl. des Particules 1, 1211 Geneva, Switzerland}

\renewcommand{\exp}{\ensuremath{\operatorname{exp}}}
\renewcommand{\epsilon}{\ensuremath{\varepsilon}}
\newcommand{\normf}{\ensuremath{\Vert f - g \Vert}}
\newcommand{\uf}{\ensuremath{U^{\rm f}_L(\bm\theta, \bm\phi, \lambda; x)}}
\newcommand{\uw}{\ensuremath{U^{\rm w}_L(\bm\theta, \bm\phi, W, \lambda; \bm x)}}
\newcommand{\uxf}{\ensuremath{U^{\rm mf}_L(\bm\theta, \bm\phi, \lambda; \bm x)}}
\newcommand{\uxfprime}{\ensuremath{U^{\rm mf}_{L^\prime}(\bm\theta^\prime, \bm\phi^\prime, \lambda^\prime; \bm x)}}
\newcommand{\ufm}{\ensuremath{U_L(\Theta, \Phi, \bm\lambda; \bm x)}}
\newcommand{\R}{\ensuremath{R_{j, k}(\bm\theta_{j, k}, \bm\phi_{j, k}, x)}}

\newtheorem{definition}{Definition}
\newtheorem{lemma}{Lemma}
\newtheorem{theorem}{Theorem}
\newtheorem{corollary}{Corollary}

\let\section\section

\crefname{paragraph}{Sec.}{Sec.}

\usepackage{titlesec}
\titleformat{\section}[runin]
       {\itshape}
       {\thesection.}% the label and number
       {0.5em}% 
       {\itshape}[ --- ]% 
\titlespacing*{\section}{10pt}{.5\baselineskip}{.2\baselineskip}

\begin{document}

\title{
Universal approximation of continuous functions with minimal quantum circuits}
\author{Adri\'an P\'erez-Salinas}
\affiliation{\aqa}
\affiliation{\lorentz}
\author{Mahtab Yaghubi Rad}
\affiliation{\aqa}
\affiliation{\liacs}
\author{Alice Barthe}
\affiliation{\aqa}
\affiliation{\cern}
\author{Vedran Dunjko}
\affiliation{\aqa}
\affiliation{\liacs}
\begin{abstract}
The conventional paradigm of quantum computing is discrete: it utilizes discrete sets of gates to realize bitstring-to-bitstring mappings, some of them arguably intractable for classical computers.  In parameterized quantum approaches, the input becomes continuous and the output represents real-valued functions. 
While the universality of discrete quantum computers is well understood, basic questions remained open in the continuous case.
We focus on universality on multivariate functions. Current approaches require either a number of qubits scaling linearly with the dimension of the input for fixed encodings,  or a tunable encoding procedure in single-qubit circuits. 
The question of whether universality can be reached with a fixed encoding and sub-linearly many qubits remained open for the last five years.
In this paper, we answer this question in the affirmative for arbitrary multivariate functions. We provide two methods: (i) a single-qubit circuit where each coordinate of the arguments to the function to represent is input independently, and (ii) a multi-qubit approach where all coordinates are input in one step, with number of qubits scaling logarithmically with the dimension of the argument of the function of interest.
We view the first result of inherent and fundamental interest, whereas the second result opens the path towards representing functions whose arguments are densely encoded in a unitary operation, possibly encoding for instance quantum processes. 
\end{abstract}

\maketitle  

\section{Introduction } 
Quantum computing extends the paradigm of classical circuits by adding new elements to the set of available logical operations, as allowed by the principles of quantum mechanics. 
The conventional paradigm of quantum computing is discrete and relies on a discrete set of gates to realize probabilistic bitstring-to-bitstring mappings. 
Some of these mappings are believed to be intractable for classical computers~\cite{shor1997polynomialtime}.  
Motivated by near-term quantum computing and the limitations of their experimental realizations, recently there has been substantial interest in settings where quantum computers are used to realize 
\textit{continuous-valued functions depending on continuous parameters}, which we refer to as variational quantum computing.
Such variational approaches may be utilized in a plethora of contexts, including quantum optimization~\cite{peruzzo2014variational, farhi2014quantum, cervera-lierta2021metavariational} or quantum machine learning \cite{havlicek2019supervised, schuld2019quantum, mitarai2018quantum, perez-salinas2020data}, among others. 
In all of these cases, the question of the set of functions such a family can represent  -- i.e. the universality of such a function family -- is central.

For discrete computation, 
universality is well understood, both in classical and quantum cases. 
Both classical and quantum computing provide universality in terms of the capacity to represent (or compute) arbitrary boolean functions\footnote{In the quantum case, we also have the universality of (approximately) representing arbitrary unitary transformations, which is also guaranteed for any so-called universal gate set, but this is tangential to our discussion}.
In the classical case, Boolean universality can be achieved 
in constant-width settings \cite{sheffer1913set, barrington1989boundedwidth}. 
Quantum computing can surprisingly do better by representing arbitrary boolean functions over $n$ bits using just a single-qubit wire~\cite{cosentino2013dequantizing}. 
For the continuous classical case, much is known about the representation power of many models such as neural networks \cite{cybenko1989approximation, hornik1991approximation}, or via e.g. the Kolmogorov-Arnold representation theorems \cite{arnold2009representation}, but for the quantum analogous fundamental open problems persisted.

Various methods to represent continuous functions with parametrized quantum circuits exist, mainly depending on how the arguments $\bm x$ of the functions are encoded \cite{perez-salinas2020data, yu2024nonasymptotic, schuld2021effect, motlagh2024generalized, silva2022fourierbased}. We consider these encoding gates to be interleaved with parameter-dependent gates to construct a larger quantum circuit $U(\bm\theta, \bm x)$ \footnote{In this work, we will use the words \textit{parameters} and \textit{arguments} for making the difference between $\bm\theta$ and $\bm x$ explicit, respectively.}. The type of encoding has direct impact in the number of parameters needed to guarantee universality in the functions that a quantum circuit can represent. For $m$-variate functions, fixed encoding gates require the number of tunable parameters to scale as $\mathcal O(\exp{(m)})$ and the number of qubits as $\mathcal O(m)$ \cite{yu2024nonasymptotic, schuld2021effect, casas2023multidimensional}. In contrast, single-qubit circuits suffice to guarantee universality if the coordinates of $\bm x$ admit pre-processing through arbitrary linear combinations before being input to the encoding gates \cite{perez-salinas2021one}. The existence of universality guarantees in the scenario where $\bm x$ is densely encoded, i. e. maximally using the size of the Hilbert space through a fixed $\mathcal O(\log m)$-qubit encoding gate, remains an open question. Note that pre-processing is not available in this scenario due to the large number of coordinates in $\bm x$.

\begin{table*}[t] 
    \centering
    \begin{tabular}{||>{\centering}p{3cm} |>{\centering}m{3cm}|>{\centering}m{2cm}|>{\centering}m{2cm}|>{\centering}m{2cm}|>{\centering}m{2cm}|>{\centering}m{2cm}|| c }\cline{1-7}
      & Reference & Pre-process  & \# qubits & \# parameters & Functions & Norm &  \\[1ex] \cline{1-7}
       \multirow{2}{4em}{Foundational \\ results} & Fourier \cite{dirichlet2008convergence},\Cref{eq.fourier} & \xmark & - & $\mathcal O(L^m)$ & $L_2$-integrable & $\Vert \cdot \Vert_2$ & \\ \cline{2-7}
         & UAT \cite{cybenko1989approximation}, \Cref{eq.uat},  & \cmark & - & $\mathcal O(L m)$ & \centering Continuous &  $\Vert \cdot \Vert_\infty$ & \\ \cline{1-7}
        \multirow{5}{4em}{Quantum \\ Circuits} & QSP (for $m = 1$) \cite{motlagh2024generalized}, \Cref{th.dru_fixed} and  Yu\&Chen's PQC \cite{yu2024nonasymptotic}, \Cref{th.dru_multi_fixed}&  \xmark  & $ \mathcal O(m)$ & $\mathcal O(L^m)$ &  Continuous & $\Vert \cdot \Vert_\infty$ & \\ \cline{2-7}
         & Data re-uploading \cite{perez-salinas2021one}, \Cref{th.dru_tunable} & \cmark  & 1 & $\mathcal O(Lm)$ &  Continuous & $\Vert \cdot \Vert_\infty$ & \\ \cline{2-7} 
          & This work, \Cref{th.multivariate_fixed} & \centering \xmark  & 1 & $\mathcal O(Lm)$ & Continuous & $\Vert \cdot \Vert_\infty$ & \\ \cline{2-7}
         & This work, \Cref{th.multidimension} & \xmark  & $\mathcal O(\log m)$ & $\mathcal O(Lm)$ &  Continuous & $\Vert \cdot \Vert_\infty$ & \\ \cline{1-7} 
    \end{tabular}
    \caption{Summary of existing results for universality of different function families and quantum circuits. The results in this work improve state-of-the-art requirements by exponential scalings. }
    \label{tab:results}
\end{table*}

In this work, we address the minimal qubit requirements to represent multivariate arbitrary functions with fixed encoding gates, and obtain universality guarantees for $\mathcal O(\log m)$-qubit fixed encoding gates, and provide a comprehensive comparison to previous results; see \Cref{tab:results}. We consider two scenarios. If each coordinate of $\bm x$ is input independently from others, single-qubit quantum circuits are universal for multivariate functions of arbitrary dimension. In the dense scenario where a $m$-dimensional $\bm x$ is densely encoded in a fixed gate, universality holds for only $\mathcal O(\log m)$-qubit circuits. These findings bridge the gap between the existing results and further reduce the qubit requirements from the previous $\mathcal O(m)$. 

\section{Background on universality}\label{sec.background}

Parameterizing arbitrary functions allows us to approximate them through controllable function families. 
The expressivity of these function families quantifies the range of implementable maps. Maximal expressivity is commonly referred to as universality, guaranteeing that arbitrary functions can be approximated. 
The formal definition of universality is as follows. 
\begin{definition}[Universality]\label{def.universality}
    Let $\mathcal G_L = \{ g\}$ be a sequence of function families, $\mathcal G_L \subseteq \mathcal G_{L + 1}$. Let $\mathcal F = \{ f\}$ be a reference function family, and $g, f: \mathcal X \subset \mathbb R^m \rightarrow \mathbb C$. Let $p$ define a function norm as 
    \begin{equation}
    \Vert f \Vert_p = \left(\int_{\bm x \in \mathcal X} \vert f(\bm x)\vert^p d\bm x\right)^{1/p}.
\end{equation} 
The sequence $\mathcal G_L$ is universal with respect to $\mathcal F$ for the $p$-norm if
\begin{equation}
    \forall f \in \mathcal F, \forall \epsilon > 0, \quad \exists L, g\in \mathcal G_L\quad  {\rm s. t. }\quad \normf_p \leq \epsilon.
\end{equation}
\end{definition}

Next, we cover some universality results. We remark that these results will be illustrative for our findings, but the list is by no means an exhaustive review. 
The first result is the known Fourier's theorem, relying on functions of the form
\begin{equation}\label{eq.fourier}
    g^F_L(\bm x) =  \sum_{\substack{\bm n \\ \Vert \bm n \Vert_1 \leq L }}c_{\bm n} e^{i 2\pi \bm n \cdot \bm x}, 
\end{equation}
for $\bm x \in \mathbb R^m, \bm n \in \mathbb Z^m$. These functions are universal in the $\Vert \cdot \Vert_2$ norm for square-integrable periodic functions. This result makes apparent that universality for functions of the form of $g_L^F(\bm x)$ demands $\mathcal O(L^m)$ degrees of freedom. 
Alternatively, the universal approximation theorem (UAT) guarantees universality for functions of the form 
\begin{equation}\label{eq.uat}
    g^{\rm UAT}_L(x) = \sum_{n = 1}^L \alpha_n  \exp{\left(i (\bm w_n \cdot \bm x + \phi_n)\right)},
\end{equation}
with $\bm w_n \in \mathbb R^m, \alpha_n, \phi_n \in \mathbb R$, for continuous functions in the $\Vert \cdot \Vert_\infty$ norm. This theorem requires only $\mathcal O(Lm)$ degrees of freedom. Detailed statements of both theorems are available in \Cref{th.fourier} and \Cref{th.uat} respectively. 

The differences between the universality properties of both function families are rooted in the presence of tunable parameters $\bm w_n$. In short, \Cref{eq.fourier} decomposes a function into a chosen basis of the space of functions. However, \Cref{eq.uat} decomposes a fucntion into terms with non-zero overlaps with all the (infinitely many) elements of any basis for the function space. As a consequence, the function space is densely covered. We refer the interested reader to \cite{cybenko1989approximation} and the proof of UAT for an in-depth discussion. The constraints on $\Vert \cdot \Vert_{\{2, \infty\}}$ norms or continuous vs. square-integrable can be overcome by choosing different function families, for instance Fejér's trigonometric or Bernstein's polynomial approximations \cite{turan1970fonctions, arato1957probabilistic}. Another component is the periodicity of the functions, which can be overcome by symmetrizing the function and appropriately adapting the domain~\cite{manzano2023parametrized}.

\begin{figure*}
    \includegraphics[width = \linewidth]{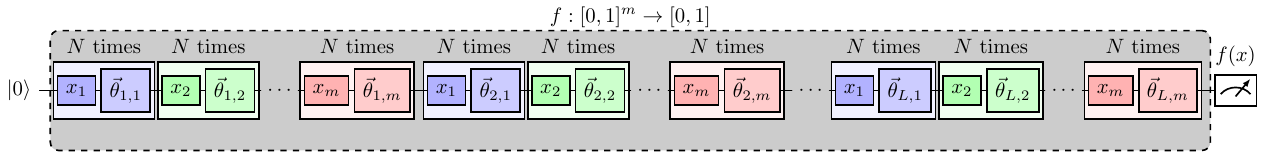}
\end{figure*}

\begin{figure}
    \includegraphics[width = \linewidth]{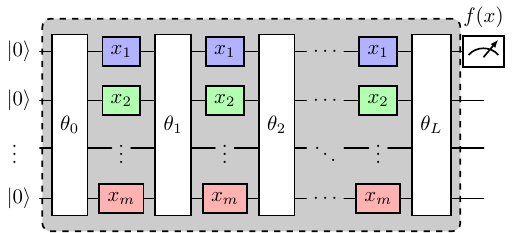}
    \begin{textblock}{0.5}(3, -2.6)
        \Huge$\Updownarrow$
    \end{textblock}
    \caption{Quantum circuits with parameters and fixed encoding gates can represent multivariate continuous functions with multiple qubits. In the figures $x_i$ is a shorthand for $\exp(i \sigma_z x_i)$ and $\theta_{i, j}$ for an arbitrary parameterized gate. The top figure corresponds to our main result, in \Cref{th.multivariate_fixed}, while bottom figure is a common approach in re-uploading models, e. g. \cite{schuld2021effect, yu2024nonasymptotic}. This figure serves as a summary for the paper: we show that single-qubit quantum circuits with fixed encoding gates retain universal representation capabilities for multivariate functions.}\label{fig.circuits}
\end{figure}

\section{Universality in parameterized quantum circuits}\label{sec.universality_quantum}

In this work, we consider quantum circuits to encode a given function. These circuits combine arguments $\bm x$ and tunable unitary gates in a global operation $U_L(\bm\theta, \bm x)$, where $L$ is a parameter to scale the size and depth of the circuit. We define the hypothesis functions given by the quantum model as 
\begin{equation}
    h_{L}(\bm x) = \bra 0 U_L(\bm\theta, \bm x) \ket 0,
\end{equation}
Notice that $h_{L}$ represents a complex number which cannot be retrieved through measurements. The choice above will be used in this manuscript as a mathematical formulation. The reduction to a measurable quantity can be achieved, for instance, by taking the modulus of $h_L(\bm x)$, among other choices. 
We call a family of quantum circuits universal if the sequence of function families $\mathcal H_L = \{h_{L}\}$ is universal, under the conditions of \Cref{def.universality}. 

We cover now relevant results on universality for quantum circuits with tunable parameters. The arguments $\bm x$ are input to the quantum circuit through encoding gates, which in certain cases admit tunable parameters. Analogously to the discussion in \Cref{sec.background}, the presence of tunable encodings allows for the distinction of two cases.

First, we focus on fixed encodings, that is, cases in which $\bm x$ is introduced through queries to unitary operations which \textit{only} depend on the data. For univariate functions ($m = 1$), it is possible to use a method based on QSP \cite{low2017optimal} to reach arbitrary polynomials up to degree $L$ with single-qubit circuits and $\mathcal O(L)$ degrees of freedom, see \Cref{th.dru_fixed} \cite{motlagh2024generalized}. Depending on the polynomials to be encoded, this circuit provides universality with different conditions. For multivariate functions of arbitrary $m$, there exist constructions for arbitrary $m$-dimensional (Bernstein's) polynomials to degree $L$ with $\mathcal O(m \log L)$ qubits and $\mathcal O(m L^m)$ parameters, see \Cref{th.dru_multi_fixed} \cite{yu2024nonasymptotic}. This circuit family provides universality for continuous functions.
If we drop the condition for fixed encoding gates, it is possible to achieve universality for multivariate functions with single-qubit circuits and $\mathcal O(Lm)$ parameters, for continuous functions in the $\Vert \cdot \Vert_\infty$, see \Cref{th.dru_tunable} \cite{perez-salinas2021one}. The different conditions for the considered scenarios are analogous to those discussed in \Cref{sec.background}. The observations there given can be easily transported to results on quantum circuits. 

\begin{figure}[t!]
    \centering
    \hrule\vskip3mm
    \includegraphics[width=0.425\linewidth]{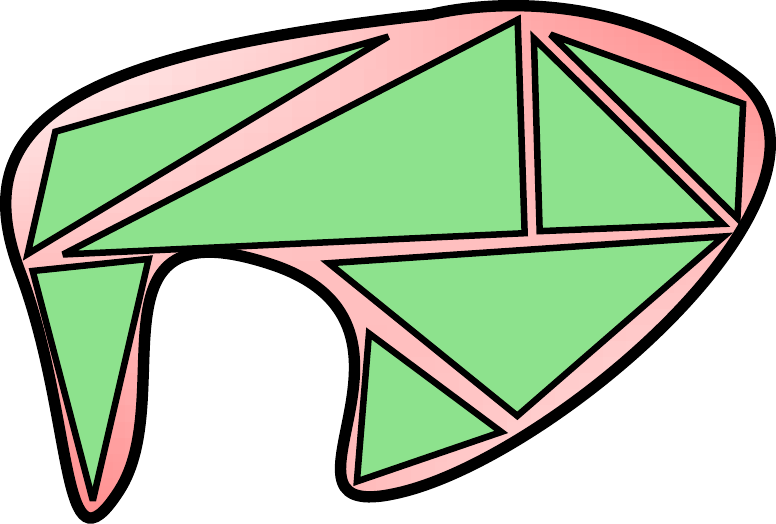}
    \raisebox{1cm}{\huge $\Rightarrow$}
    \includegraphics[width=0.425\linewidth]{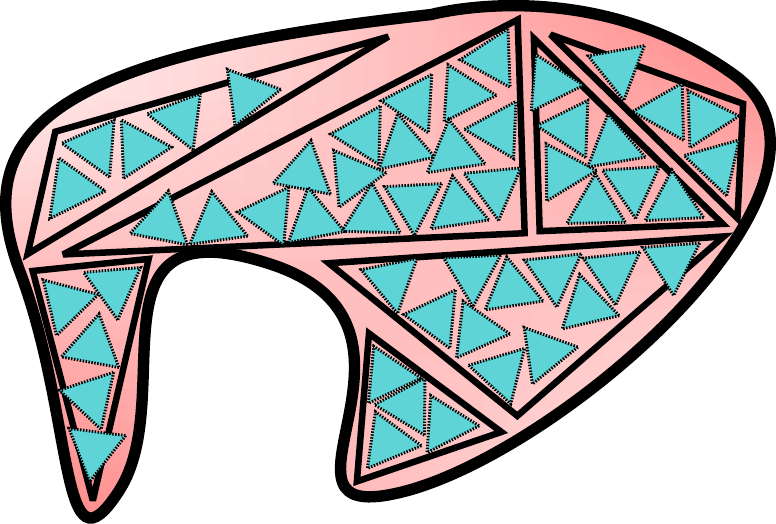}
    \caption{High level illustration for the basic concepts of the proof for universality. Consider the curved blob to be a representation of a set of functions. Triangles are components specified by different gates in the circuit. The re-uploading model with tunable parameters (left) is represented by triangles with tunable sizes and shapes, while fixed encoding gates (right) are represented by fixed-shaped triangles, but tunable in size. We can approximate the large tunable triangles with small fixed-shaped triangles. In the limit, any curved blob can be approximated by many small triangles. }
    \label{fig.reuploading_proof}
\end{figure}

To question the impact of quantum circuits, it is interesting to consider how the arguments $\bm x$ are encoded. The only operation that allows such encoding is querying a $\bm x$-dependent unitary operation. In the context of quantum circuits, one can then argue that having fixed encoding gates implies that $\mathcal O(\exp(m))$ parameters are required to achieve universality. This is a direct consequence of the models considered for fixed encodings, where each coordinate of the arguments $x_i, i \in \{1, 2, \cdots, m\}$ is independently input to the quantum circuits, for instance through unitaries composed of tensor products. On the other hand, tunable encodings seem to be required to achieve universality with $\mathcal O(m)$ many parameters. 
The trade-off between controllability of the encoding or number of degrees of freedom in principle hinders multidimensional analogous to QSP, that is, scenarios where $m$ coordinates are simultaneously and densely encoded to a quantum circuit without ancillary qubits. Surprisingly, we show in the next section that there exists a construction to circumvent this trade-off.

\section{Results}\label{sec.universality}

Universality for multivariate functions with fixed encoding gates is possible.    
We begin by defining a multivariate single-qubit re-uploading model with fixed encoding gates. 
\begin{definition}[Multivariate fixed re-uploading]\label{def.dru_multi_fixed}
A single-qubit multivariate data re-uploading circuit with fixed encoding gates is given by
\begin{multline}\label{eq.dru_multi_fixed}
        \uxf = \\ \left(\prod_{j = 1}^L \prod_{k = 1}^m \prod_{n = 1}^N e^{i \theta_{j, k, n}\hat\sigma_z} e^{i \phi_{j, k, n}\hat\sigma_y} e^{i \pi x_{k}\hat\sigma_z} \right) \\  e^{i \theta_{0}\hat\sigma_z} e^{i \phi_{0}\hat\sigma_y} e^{i \lambda\hat\sigma_z},
\end{multline}
where $\bm\theta, \bm\phi, \lambda$ are real and tunable. 
\end{definition}

The first main result of this manuscript holds as follows.

\begin{theorem}\label{th.multivariate_fixed}
    Consider the model $\uxf$ from~\Cref{eq.dru_multi_fixed}. The sequence of output function families $\mathcal H = \{h_{L}^{\rm f} \}$, with  
    \begin{equation}
        h_{L}(\bm x) = \bra 0 \uxf \ket 0
    \end{equation}
    is  universal with respect to multivariate continuous functions $f : [0, 1]^m \rightarrow \mathbb C$ with the constraint $\vert f(\bm x)\vert^2 \leq 1$, in the $\norm{\cdot}_\infty$ norm. 
\end{theorem}

We provide a sketch of the proof, with details given in \Cref{app.multivariate_fixed} and the references therein.
As discussed in the previous section, the key to universality with a small number of parameters is to construct a family of functions whose elements overlap with all (infinitely many) elements of any basis of the function space. For instance, Fourier expansions are restricted to functions in the basis $\exp(i n x)$ with $n \in \mathbb Z$. In contrast, $\exp(i \alpha x)$ with $\alpha \in \mathbb R$ overlaps with all elements of the Fourier basis simultaneously. This observation was used in \Cref{th.uat} to establish universality without relying on a specific basis. In the case of the quantum circuit, fixed inputs limit the expressivity to functions of the form $\exp(i n x)$, since encoding gates can only be applied an integer number of times. However, by quantum signal processing (QSP) \cite{motlagh2024generalized}, we can approximate the univariate function $\exp(i \hat\sigma_z w x)$ to arbitrary precision (see \Cref{le.one_gate}). The next step is to show that these univariate approximations can be composed to approximate a multivariate gate of the form $\exp(i \hat\sigma_z \bm w \cdot \bm x)$. Finally, applying the triangle inequality shows that if such circuits can approximate any continuous function, and these circuits can themselves be approximated with fixed encoding gates, then the construction in \Cref{def.dru_multi_fixed} achieves universal approximation.
A graphical summary of this reasoning is shown in \Cref{fig.reuploading_proof}.

As discussed in the previous section, the key element for attaining universality with a small number of parameters is to create a family of functions whose elements overlap with all (infinite) elements in any basis of the space of functions. As an example, Fourier expansions are limited to representing a function in the basis $\exp{(i n x)}, n \in \mathbb Z$. However, $\exp{(i \alpha x)}, \alpha \in \mathbb R$ does overlap with all elements of the Fourier basis simultaneously, and this fact was used in \Cref{th.uat} to prove universality without relying on basis of the space of functions. Our quantum circuit mimics this idea. Fixed inputs limits our expressivity to functions $\exp{(i nx)}$ since we can only apply the encoding gates an integer number of times. Note however that we can approximate $\exp{(i \hat\sigma_z w x)}$, a univariate function, to arbitrary precision using QSP \cite{motlagh2024generalized}, see \Cref{le.one_gate}. Next, we need to show that the aforementioned approximations can be stacked to obtain another approximation for a multivariate gate of the form $\exp{(i \hat\sigma_z \bm w \cdot \bm x)}$. The last step is to apply a triangle inequality. If such circuit can approximate any continuous function, and this circuit can be approximated with fixed encoding gates, then the circuit in \Cref{def.dru_multi_fixed} can approximate any function to arbitrary precision. See \Cref{fig.reuploading_proof} for a graphical interpretation of this proof. 

\Cref{th.multivariate_fixed} shows that it is possible to approximate arbitrary continuous multivariate functions with single-qubit circuits and fixed encoding gates of the form $\exp{(i \hat\sigma_z \pi x_k)}$. The circuits to achieve this require to query the $m$ different encoding gates, (one per coordinate) on demand. This result implies a reduction in the requirements of number of qubits from linear in $m$ to just one. This result is analogous to Barrington's theorem for classical circuits \cite{barrington1989boundedwidth, cosentino2013dequantizing}.

For completeness, we can extend the obtained result to a more general theorem of functional analysis, independent from quantum circuits. 
\begin{corollary}\label{cor.analysis}
    Let $\mathcal G_{L, N} = \{ g_{L, N}\}$ be the sequence of sets of functions $g_N: [0, 1]^m \rightarrow \mathbb C$ of the form
    \begin{equation}
        g_{L, N}(\bm x) = \sum_{j = 1}^L \gamma_j \prod_{k = 1}^m \sum_{n = -N}^N c_{j, k, n}e^{i \pi n x_k},
    \end{equation}
    for $\bm x \in [0, 1]^m, \bm w_n \in \mathbb R^m, \alpha_n, \phi_n \in \mathbb R$. Then, $\mathcal G_{L, N}$ is universal 
    with respect to the set of all continuous functions $f:[0, 1]^m \rightarrow \mathbb C$, in the norm $\norm{\cdot}_\infty$.
\end{corollary}
The proof can be found in~\Cref{app.analysis}.

For the second main result of this manuscript we consider a generalization of the univariate encoding $e^{i w x\hat\sigma_z}$ to a multivariate case. 
The motivation to study these cases stems from considering the arguments input to the system as the query of a unitary gate carrying all relevant information. We consider diagonal encoding, which needs $\mathcal O(\log m)$ qubits to carry $\mathcal O(m)$ real numbers. 
Specifically, we consider the case where the $m$-dimensional input is given by a matrix of the form
\begin{equation}\label{eq.multi}
    V_m(\bm x) = \sum_{k = 1}^m \ketbra{k}{k} \otimes e^{i \pi x_k\hat\sigma_z/2}. 
\end{equation}
This scenario contrasts with other approaches where the arguments are carried by a tensor product of $m$ matrices of size $2 \times 2$ \cite{schuld2021effect}. In \Cref{app.extension}, we consider an even more restricted additional case when the encoding matrix contains each coordinate of the input as a diagonal element.

\begin{definition}[Re-uploading: fixed encoding gates and arbitrary dimensions]\label{def.dru_fixed_large}
A multi-qubit multivariate circuit with fixed encoding gates is given by
\begin{multline}\label{eq.dru_fixed_large}
    \ufm = \\ 
    \left(\prod_{j = 1}^L R(\bm\theta_j, \bm\phi_j) V_m(\bm x) \right)
    R(\bm\theta_0, \bm\phi_0) V_m(\bm \lambda) ,
\end{multline}
where $R(\bm\theta, \bm\phi)$ are the specifications through Euler angles of arbitrary $\mathcal{SU}(2m)$ matrices. 
\end{definition}

Our quantum circuit only includes $V_m(x)$ and not any controlled version of it. Thus, we fix the relative phase with extra dimensions, ensuring the unitary is an element of in $\mathcal{SU}(2m)$. However, this choice does not hinder the generality of our arguments. Any change in the relative phase can be given by an affine transformation of $\bm x$. 
Following the rationale of QSP (\Cref{th.dru_fixed}), we note that $V_m(\bm x)$ plays the role specific angles in the Euler decomposition of $\mathcal{SU}(2m)$. The extension to Euler decomposition of arbitrary dimensions~\cite{tilma2002generalized} shows that $R(\bm\theta, \bm\phi)$ is specifies any unitary matrix in $\mathcal{SU}(2m)$, except for the relative phases among columns, fixed by $\bm x$. We refer the reader to~\Cref{app.euler} for a detailed description. 

We can now state the final result. 

\begin{theorem}\label{th.multidimension}
Let $\ufm$ be the gate defined in~\Cref{def.dru_fixed_large}, with $\mathcal O(\log m)$ many qubits. The sequence of output function families $\mathcal H = \{h_{L} \}$, with  
    \begin{equation}
        h_L(\bm x) = \bra 0 \ufm \ket 0
    \end{equation}
    is universal for all continuous complex functions $f(\bm x), \norm{f(\bm x)}_\infty \leq 1$, for $\bm x \in [0, 1]^m$,  with respect to the $\norm{\cdot}_\infty$ norm. 
\end{theorem}

The proof can be found in~\Cref{app.multidimension}. This model has a constant overhead in depth as compared to the single-qubit case from \Cref{th.multivariate_fixed}. The sketch of the proof is as follows. There exists a choice of $\Theta, \Phi$ such that the concatenation of layers of the form $R(\bm\theta_j, \bm\phi_j) V_m(\bm x)$ provides a matrix in a block form, such that the first block is $e^{i \pi x_k\hat\sigma_z}$, for any $k$, and the rest is the identity matrix. This transforms a multidimensional problem into a concatenation similar to~\Cref{def.dru_multi_fixed}. Application of results from \Cref{th.multivariate_fixed} provides universality. 
In \Cref{app.extension}, we consider an even more condensed additional case when the encoding matrix contains each coordinate of the input as a diagonal element.

The results here described open an avenue towards representing functions in the case where the arguments are accessible via query operations and exponentially many coordinates are simultaneously available, thus more explicitly utilizing the exponential size of the Hilbert space. 
An example of such operations on demand are time evolution dynamics, interleaved with tunable gates. Other existing methodologies encode argument coordinates individually, for example, by means of encoding gates composed of tensor products \cite{schuld2021effect, yu2024nonasymptotic}, thus severely limiting for instance the applicability of learning processes to scenarios with classical data. In addition to this advantage, our construction requires a qubit scaling of $\mathcal O(\log m)$, improving previous linear scalings. Consequently, the requirement for the number of parameters required to specify these models is reduced from $\exp{(\mathcal{O}(m))}$ to just $\mathcal{O}({\rm poly}(m))$. One can interpret this as an efficient generalization of QSP in which the signal is densely encoded in the form of independent quantities inside a unitary gate of dimension $2^m$. These properties are minimal requirements for the existence of quantum advantages in variational learning problems, to be further investigated. 

\section{Conclusions}\label{sec.conclusions}
In this paper, we show that it is possible to encode any continuous function of arbitrary dimensions using single-qubit quantum circuits with fixed encoding gates. This finding extends existing results of universality for multivariate functions with fixed encodings on multi-qubit, and multivariate functions with tunable encodings in single-qubit circuits.

As a second result, we show that universality on multivariate functions is also attainable if the arguments are input by a diagonal unitary matrix in $\Theta(\log m)$ qubits, where $m$ is the dimension of the function to represent. This finding opens a path towards $m$-dimensional generalizations of QSP where the argument $\bm x$ is given as the entries of a unitary matrices, for instance, time evolution, and opens the path for broader quantum machine learning algorithms. 

The question of whether a trade-off between tunability of encoding and the number of qubits is needed is now closed. The results here depicted can be interpreted as analogous to width-depth trade-offs in boolean classical and quantum computing. Our results support universality by showing the existence of circuits with arbitrarily accurate approximations to functions. The performance of those models in practice, for instance related to trainability or (non-)dequantizability is a valuable future research direction. The extension of QSP techniques to multivariate functions with a reduced number of parameters paves the way towards machine learning approaches which take black-box quantum operations as inputs. 

\acknowledgments

The authors would like to thank members of aQa Leiden for fruitful discussions. 
This work was supported by the Dutch National Growth Fund (NGF), as part of the Quantum Delta NL programme.
This publication is part of the project Divide \& Quantum (with project number 1389.20.241) of the research programme NWA-ORC which is (partly) financed by the Dutch Research Council (NWO)
This work was also partially supported by the Dutch Research Council (NWO/OCW), as part of the Quantum Software Consortium programme (project number 024.003.03), and co-funded by the European Union (ERC CoG, BeMAIQuantum, 101124342).
AB is supported by CERN through the CERN Quantum Technology Initiative. AB is supported by the quantum computing for earth observation (QC4EO) initiative of ESA $\phi$-lab, partially funded under contract 4000135723/21/I-DT-lr, in the FutureEO program.

\bibliography{references}

\onecolumngrid

\newpage

\appendix

\begin{center}
    \Large\bf Appendix
\end{center}

\subsection{Relevant theorems on universality}\label{app.universality}

\begin{theorem}[Fourier theorem]~\label{th.fourier}
        Let $\mathcal G^F_L = \{ g^F_L(x)\}$ be the sequence of sets of functions $g^F_L: [0, 1]^m \rightarrow \mathbb C$ of the form
    \begin{equation}
        g^F_L(\bm x) =  \sum_{\bm n}c_{\bm n} \exp{(i 2\pi \bm n \cdot \bm x)} 
    \end{equation}
    for $\bm n = (n_1, \ldots, n_m) \in \mathbb N^m, \norm{\bm n}_\infty \leq L$ and $c_{\bm n}\in \mathbb C$.  For every square-integrable function $f: [0, 1]^m \rightarrow \mathbb C, f \in \mathcal F$, the sequence $\mathcal G^F_L$ is universal with respect to $\mathcal F$ in the norm $\norm{\cdot}_2$, if the coefficients are given by 
    \begin{equation}
        c_{\bm n} = \int_{[0, 1]^m} d\bm x f(\bm x) e^{-i 2\pi \bm n \cdot \bm x}
    \end{equation}
\end{theorem}

The interpretation of this theorem is direct in the language of Hilbert spaces. We consider functions as elements of an infinite-dimensional Hilbert space. In this case, the basis of the space is the set $\{ e^{i 2\pi \bm n \cdot \bm x}\}$, with $n$ a vector of integers. The Hilbert space is equipped with the inner product
\begin{equation}
    \langle f(\bm x), g(\bm x) \rangle = \int_{[0, 1]^m} f^*(\bm x) g(\bm x) d\bm x, 
\end{equation}
and in particular
\begin{equation}
    \langle e^{i 2\pi \bm n \cdot \bm x}, e^{i 2\pi \bm l \cdot \bm x} \rangle = \prod_{k = 1}^m \delta_{n_k l_k},
\end{equation}
thus yielding an orthonormal basis. The Fourier theorem is nothing but the transformation of functions from arbitrary forms to this Hilbert space picture. 

\begin{theorem}[Universal approximation theorem, adapted from~\cite{cybenko1989approximation}]\label{th.uat}
    Let $\mathcal G^{\rm UAT}_L = \{ g^{\rm UAT}_L(x)\}$ be the set of functions $g^{\rm UAT}_L: [0, 1]^m \rightarrow \mathbb C$ of the form
    \begin{equation}
        g^{\rm UAT}_L(x) = \sum_{n = 1}^L \alpha_n \hat\sigma\left(\bm w_n \cdot \bm x + \phi_n\right),
    \end{equation}
    for $\bm x \in [0, 1]^m, \bm w_n \in \mathbb R^m, \alpha_n, \phi_n \in \mathbb R$, and $\sigma(\cdot)$ a discriminatory function. For every continuous function $f: [0, 1]^m \rightarrow \mathbb C, f \in \mathcal F$ there exists a $N$ such that $\mathcal G^{\rm UAT}_N$ is universal with respect to $\mathcal F$ in the norm $\norm{\cdot}_\infty$. Discriminatory functions are those satisfying 
    \begin{equation}
        \int_{[0, 1]^m} d\mu(\bm x)\hat\sigma(\bm w \cdot \bm x + \phi) = 0, \; \bm w \in \mathbb R^m, \phi \in \mathbb R \Leftrightarrow \mu = 0,
    \end{equation}
    for $\mu$ being Borel measures. In particular, $e^{i \ (\cdot)}$ is a discriminatory function.
\end{theorem}

The interpretation of this theorem can be done in terms of the Hilbert space discussed after~\Cref{th.fourier}. For arbitrary $\bm w, \phi$, discriminatory functions satisfy that they have non-zero overlaps with all elements in the basis of functions. For exponential functions 
\begin{align}
        e^{iwx} & = \sum_{n = -\infty}^\infty c_n(w) e^{i n 2\pi  x}, \\ 
        c_n(w) & = \frac{-i}{ ( 2\pi n - w )} \left( e^{iw} - 1\right).
    \end{align}
This property, together with the Hahn-Banach theorem, guarantees dense covering of the space of continuous functions, hence universality. We refer the reader to the original statement of the UAT in Ref.~\cite{cybenko1989approximation} for an in-depth proof.

\subsection{Relevant theorems on universality of quantum circuits}\label{app.universality_quantum}

\begin{theorem}[Re-uploading: fixed encoding gates, adapted from~\cite{silva2022fourierbased, motlagh2024generalized}]\label{th.dru_fixed}
Consider the single-qubit circuits
\begin{equation}\label{eq.dru_fixed}
    \uf = \left(\prod_{j = 1}^L e^{i\hat\sigma_z \theta_{ j}}
    e^{ i\hat\sigma_y \phi_{j}}e^{ i\hat\sigma_z 2\pi  x}\right)e^{ i\hat\sigma_z \theta_{0}}
    e^{ i\hat\sigma_y \phi_{0}} e^{ i\hat\sigma_z \lambda},
\end{equation}
    where $\sigma_{\{y, z\}}$ are Pauli matrices, and $x \in [0, 1]$. The sequence of output function families $\mathcal H_L = \{ h_L\}$, $h_{L}: [0, 1] \rightarrow \mathbb C, \vert h_{L}(\bm x) \vert \leq 1, h_{\bm\theta}(x) = \bra 0 \uf \ket 0$ is universal with respect to univariate square-integrable functions $f : [0, 1] \rightarrow \mathbb C$ with the constraint $\vert f(\bm x)\vert^2 \leq 1$ in the $\norm{\cdot}_2$ norm.
\end{theorem}
\begin{proof}
    We recall theorem 3 in~\cite{motlagh2024generalized}, formulated as follows. 
    \begin{theorem}[Generalized quantum signal processing]
    Consider the building block 
    \begin{equation}\label{eq.R}
    R(\theta, \phi, \lambda) = \begin{pmatrix}
        e^{i(\phi + \lambda)} \cos(\theta) & e^{i\phi } \sin(\theta) \\
        e^{i\lambda} \sin(\theta) & - \cos(\theta)
    \end{pmatrix}.
\end{equation}
        Then, $\forall L \in \mathbb N, \; \exists \bm \theta, \bm\phi \in \mathbb R^{L + 1}, \lambda \in \mathbb R$ such that:
        \begin{equation}
            \left(\prod_{j = 1}^L R(\theta_j, 
            \phi_j, x) \right) R(\theta_0, \phi_0, \lambda) = \begin{pmatrix}
                P(e^{ix}) & \cdot \\ Q(e^{ix}) & \cdot, 
            \end{pmatrix}
        \end{equation}
        with $P(e^{ix}), Q(e^{ix})$ being polynomials of degree $L$ subject to the constraint $\vert P(e^{ix}) \vert^2 + \vert Q(e^{ix}) \vert^2 = 1, \; \forall x$.         
    \end{theorem}
    In the same reference, corollary 5 demonstrates that $P(e^{ix})$ can be chosen arbitrarily. Additionally, theorem 6 in the same reference implies that it is possible to obtain arbitrary polynomials of the form
    \begin{equation}
        P^\prime(e^{ix}) = e^{-ikx} P(e^{ix})
    \end{equation}
    if the argument is transformed according to $x \rightarrow -x$ in the last $k$ layers. 
    
    We focus on this last result. Notice that changing the sign of $x$ is equivalent to performing an inversion, as
    \begin{equation}
        \begin{pmatrix}
            e^{-ix} & 0 \\ 0 & 1
        \end{pmatrix} = e^{-ix} X \begin{pmatrix}
            e^{ix} & 0 \\ 0 & 1
        \end{pmatrix} X.
    \end{equation}
    We identify
    \begin{align}
        \begin{pmatrix}
            e^{ix} & 0 \\ 0 & 1
        \end{pmatrix} & = 
        e^{ix/2}\begin{pmatrix}
            e^{ix/2} & 0 \\ 0 & e^{-ix/2}
        \end{pmatrix} = e^{i x / 2} e^{i x/2\hat\sigma_z}\\ 
        \begin{pmatrix}
            e^{-ix} & 0 \\ 0 & 1
        \end{pmatrix} & = 
        e^{-ix/2}\begin{pmatrix}
            e^{-ix/2} & 0 \\ 0 & e^{ix/2}.
        \end{pmatrix} = e^{- i x / 2} e^{- i x/2\hat\sigma_z}
    \end{align}
    By querying $L$ times the first operator and $L$ times the second, consecutively, we can represent any polynomial by virtue of~\cite{motlagh2024generalized}. These gates are accessible with the same query, plus the unitary transformation that can be absorbed in the gates before and after. The global phases are compensated. Thus, following this recipe, we can find arbitrary polynomials of the form
    \begin{equation}
        P(e^{ix}) = \sum_{j = -L}^L c_j e^{i \cdot j x}.
    \end{equation}
    These polynomials are Fourier series up to degree $L$, and thus this family of functions is universal as stated in~\Cref{th.fourier}. 
    
    Finally, we bridge the gap between this result and our statement in~\Cref{th.dru_fixed}. Notice that in this case $x \in [0, 2\pi]$. To maintain consistency with our result, we impose $x \in [0, 1]$ at the expense of adding the factor $2\pi$ in the encoding. The building block in~\Cref{eq.R} is decomposable as~\Cref{eq.dru_fixed} up to global phases.
\end{proof}

\begin{theorem}[Re-uploading: fixed encoding gates, multivariate, adapted from \cite{yu2024nonasymptotic}]\label{th.dru_multi_fixed}
    There exists a sequence of quantum circuits $W_L(\bm x, \bm\theta)$ with number number of qubits $\mathcal O(m \log L)$, depth $\mathcal O(m L^m \log L)$ and number of trainable parameters $\mathcal O(m L^m)$, such that the sequence of output function families $\mathcal H_L = \{ h_L\}$, $h_L(\bm x) = \bra 0 W_L(\bm x, \bm\theta)\ket 0$ is universal with respect to multivariate continuous functions $f : [0, 1]^m \rightarrow \mathbb C$ with the constraint $\vert f(\bm x)\vert^2 \leq 1$, in the $\norm{\cdot}_\infty$ norm. 
\end{theorem}

\begin{theorem}[Re-uploading: tunable encoding gates~\cite{perez-salinas2021one}]\label{th.dru_tunable}
Consider the single-qubit circuit 
\begin{equation}\label{eq.dru_tunable}
    \uw = \left(\prod_{j = 1}^L e^{ i\hat\sigma_z \theta_{j}}
    e^{ i\hat\sigma_y \phi_{j}}e^{ i\hat\sigma_z \bm w_j \cdot  \bm x} \right) 
    e^{ i\hat\sigma_z \theta_{0}}
    e^{ i\hat\sigma_y \phi_{0}}e^{ i\hat\sigma_z  \lambda },
\end{equation}
    with $\sigma_{\{y, z\}}$ being Pauli matrices, and $\bm x \in [0, 1]^m$. The sequence of output function families $\mathcal H_L = \{ h_L\}$, $h_L(x) = \bra 0 \uw \ket 0$ is universal with respect to multivariate continuous functions $f : [0, 1]^m \rightarrow \mathbb C$ with the constraint $\vert f(\bm x)\vert^2 \leq 1$ in the norm $\norm{\cdot}_\infty$. 
\end{theorem}

\subsection{Proof of \Cref{th.multivariate_fixed}}\label{app.multivariate_fixed}

The road towards proving universality is to show that circuits as given in~\Cref{eq.dru_multi_fixed} can approximate circuits of the form given in~\Cref{eq.dru_tunable}, with arbitrary precision. 

The first step is to approximate gates of the form $e^{i w x\hat\sigma_z}$, where $x$ is the input and $w$ is a tunable real weight, with circuits composed of fixed encoding gates. Inspection of~\Cref{def.dru_multi_fixed} allows us to identify 
\begin{equation}\label{eq.R_gate}
    R_{j, k}(\bm\theta_{j, k}, \bm\phi_{j, k}, x) \equiv \prod_{n = 1}^N e^{i \theta_{j, k, n}\hat\sigma_z} e^{i \phi_{j, k, n}\hat\sigma_y} e^{i \pi x_{k}\hat\sigma_z}
    \approx e^{i w_k x_k\hat\sigma_z},
\end{equation}
as a requirement for the approximation. We just need to show that there exist parameters $\bm\theta_{j, k}, \bm\phi_{j, k}$, depending on $w_k \in \mathbb R$, such that the above equation holds to arbitrary precision $\epsilon$. This is possible by means of the following result. 
\begin{lemma}\label{le.one_gate}
    For a gate $R(\bm\theta, \bm\phi, x)$ as given in~\Cref{eq.R_gate}, for any $w \in \mathbb R$ and for any $\epsilon > 0$, there exists a value $N$ and parameters $\left({\bm\theta, \bm\phi}\right)$ such that 
    \begin{equation}
        \sup_{x \in [0, 1]} \norm{R(\bm\theta, \bm\phi, x) - \exp{\left(i w x\hat\sigma_z\right)}}_F \leq \epsilon, 
    \end{equation}
    where $\norm{\cdot}_F$ is the Frobenius norm.
\end{lemma}
Notice that the Frobenius norm $\Vert \cdot \Vert_F$ is a matrix norm, and not a function norm as other norms used through the text. In this and subsequent proofs, the $\infty$-norm for functions is made explicit as $\sup_{x \in [0, 1]^m}$ to avoid confusion with matrix norms. The Frobenius norm of a matrix is defined as
\begin{equation}
\Vert A \Vert_F = \sqrt{\Tr\left( A^\dagger A \right)} 
\end{equation}

\begin{proof}
    We consider in this proof only approximations to $ w \in [-\pi/2, \pi/2]$. Without loss of generality, this implies arbitrary approximations for $w \in \mathbb R$, since the integer approximation to $w$ is attainable by just repeating the encoding layer. Notice that, by definition, $f(x) = e^{iwx}$ satisfies $\vert f(x)\vert^2 = 1$. By~\Cref{th.dru_fixed} we know that it is possible to construct arbitrary polynomials $P(e^{ix})$ of degree at most $N$ into a unitary matrix as
    \begin{equation}\label{eq.R_PQ}
        R_N(\bm\theta, \bm\phi, x) = \begin{pmatrix}
            P_N(x) & -Q_N^*(x) \\ Q_N(x) & P_N^*(x)
        \end{pmatrix} \approx \begin{pmatrix}
            e^{i w x} & 0 \\ 0 & e^{-i w x}
        \end{pmatrix}.
    \end{equation}
    In the language of~\Cref{th.dru_fixed}, $P(x)$ is a Fourier-like polynomial of $e^{i w x}$ up to degree $N$. In particular, we choose the polynomial to be a Cesàro mean of the function $e^{iwx}$. 

    \begin{definition}[Cesàro means]
    Let $f : \mathbb R \rightarrow \mathbb C$ be a continuous function with period $2\pi$, and let $g_n^F(x)$ be its $n$-term Fourier series. The Cesàro mean of order $N$ is given by
    \begin{equation}
        P_N(x) = \frac{1}{N+1}\sum_{n = -N}^N g_n^F(x), 
    \end{equation}
    or equivalently
    \begin{equation}
        P_N(x) = \sum_{n = -N}^N \frac{N + 1 - \vert n \vert}{N + 1} c_n e^{in\pi x}. 
    \end{equation}
    \end{definition}
    
    This choice of $P_N(x)$ implies several convenient properties. First, $P_N(x)$ can be understood as the convolution of the function $e^{iwx}$ with the Féjer Kernel \cite{turan1970fonctions}
    \begin{equation}
        K_N(x) = \frac{1}{N + 1}\sum_{n = 0}^{N} \sum_{k = -n}^n e^{i k x}. 
    \end{equation}
    which satisfies
    \begin{equation}
        \int_x dx K_N(x) = 1.
    \end{equation}
    Thus
    \begin{equation}
        \Vert P_N \Vert_\infty = \sup_{x \in [0, 1]}\left\vert \int dt e^{iw(x - t)} K_N(t) dt \right\vert \leq \sup_{x \in [0, 1]} \vert e^{iwx} \vert \left\vert \int dt K_N(t) dt\right\vert = 1. 
    \end{equation}
    Therefore, we can implement this function within a unitary operation. 

    Second, we can recall Fejér's theorem \cite{turan1970fonctions}. 

    \begin{theorem}[Fejér's theorem, \cite{turan1970fonctions}]\label{th.fejer}
    Let $f : \mathbb R \rightarrow \mathbb C$ be a continuous function with period $2\pi$, and let $P_N(x)$ be its $N$-term Cesàro mean. Then $P_N$ converges uniformly to $f$ as $N$ increases, that is for every $\epsilon > 0 $ there exists  $N$ satisfying
    \begin{equation}
        \sup_{x \in [0, 1]}\vert f(x) - P_N(x) \vert_\infty \leq \epsilon.
    \end{equation}
    \end{theorem}
Choosing $P_N(x)$ as the corresponding Cesàro mean, we choose $\bm\theta_{j, k}, \bm\phi_{j, k}$ such that $\R$ implements $P_N(x)$ and $Q_N(x)$ as in \Cref{eq.R_PQ}. With this choice, we can now write the Frobenius norm of the difference matrix as
    \begin{equation}
        \norm{
        \R - e^{i w x\hat\sigma_z}
        }_F = \sqrt 2 \sqrt{\vert P_N(x) - e^{iwx}\vert^2 + \vert Q_N(x)\vert^2}.
    \end{equation}
    The Frobenius norm can be bounded as follows. We use first the triangle inequality and convexity of functions to reach 
    \begin{equation}
        \vert e^{iwx} - P_N(x)\vert \geq 1 - \vert P_N(x) \vert \geq \frac{1 - \vert P_N(x)\vert^2}{2} = \frac{\vert Q_N(x) \vert^2}{2}, 
    \end{equation}
    hence
    \begin{equation}
        \norm{
        \R - e^{i w x\hat\sigma_z}
        }_F \leq \sqrt{2} \sqrt{\vert e^{iwx} - P_N(x)\vert^2 + 2 \vert e^{iwx} - P_N(x)\vert}.
    \end{equation}
    
    By virtue of Fejér's theorem, we can make the supremum norm of this function arbitrarily small, as long as $e^{iwx}$ satisfies the continuity assumption. We enforce this constraint by finding the Cesàro means of an auxiliary related function, following the techniques in~\cite{manzano2023parametrized}. We define the period of this auxiliary function as
    \begin{equation}\label{eq.aux_function}
        a(x) = \left\{\begin{matrix}
            e^{iwx}& \qquad & 0\leq x \leq 1 \\
            e^{iw(2 - x)}& \qquad & 1 < x < 2
        \end{matrix} \right. ,
    \end{equation}
    and compute the Cesàro means of this function. This function is equivalent to $e^{iwx}$ in the domain $x \in [0, 1]$, and it also fulfils all the requirements of Fejér's theorem. We just need to ensure that our model is capable of expressing polynomials of $e^{i n\pi x}$. This follows immediately from the definition of $\R$, which has $e^{i \pi x\hat\sigma_z}$ in it. Thus, we can find a gate $\R$ satisfying that for all $\epsilon > 0$ there exists a $N$ such that 
    \begin{equation}
        \sup_{x \in [0, 1]} \left\Vert \R - e^{iwx\hat\sigma_z} \right\Vert_F \leq \epsilon,
    \end{equation}
    and concludes the proof. 
\end{proof}

We can also bound the errors in \Cref{le.one_gate}.

\begin{corollary}\label{cor.errors}
The circuit $R(\bm\theta, \bm\phi, x)$ from \Cref{le.one_gate} can be realized using $N$ fixed-encoding gates, with
    \begin{equation}
    N \in \Tilde{\mathcal O}\left(w + \epsilon^{-2} \right). 
    \end{equation}
\end{corollary}

\begin{proof}
    We begin with the function defined in~\Cref{eq.aux_function}. This function is continuous and periodic. Its Fourier series is given by the sets of coefficients
\begin{equation}
    c_n = \int_0^1 e^{i(w-n)x} dx + e^{i 2 w}\int_1^2 e^{-i(w+n)x} = -i \frac{w \left( (-1)^n e^{i w} - 1\right)}{w^2 - (n \pi)^2}.
\end{equation}
We bound now $\sup_{x \in [0, 1]}\left\vert P_N(x) - e^{iwx} \right\vert$. The definition of $P_N(x)$ from the proof of \Cref{le.one_gate} implies
\begin{equation}
    P_N(x) - e^{iwx} = \sum_{n = -N}^N c_n \frac{- \vert n \vert }{N + 1} e^{i n\pi x} - \sum_{\vert n \vert  = N + 1}^\infty c_n  e^{i n\pi x}, 
\end{equation}
hence by the triangle inequality
\begin{equation}\label{eq.52}
    \sup_{x \in [0, 1]}\vert P_N(x) - e^{iwx} \vert \leq \frac{2}{N + 1} \left\vert \sum_{n = 1}^N n c_n \right\vert + 2 \left\vert \sum_{n = N+1}^\infty c_n \right\vert.
\end{equation}
For this proof, we consider that $\vert w \vert \leq \pi / 2 $. The reason is that we can obtain $K \pi \approx w$ exactly with $K$ gates, and from this point on it is only needed to approximate the remainder $\vert w \vert \leq \pi / 2$ using Fejér's theorem. We focus on each term individually. The first term is a 1-norm of the vector defined by $n c_n$. Thus, 
\begin{equation}
    \left\vert \sum_{n = 1}^N n c_n \right\vert \leq  \sum_{n = 1}^N n \left\vert c_n \right\vert \leq \frac{2}{\pi}\left( \sum_{n = 1}^N \frac{n}{n^2 - \frac{1}{4}} \right) \leq \frac{2}{\pi} \left( \frac{4}{3} + \int_{1}^N dx \frac{x}{x^2 - \frac{1}{4}}\right) = \frac{2}{\pi} \left( \frac{4}{3} + \frac{1}{2} \log\left( \frac{4N^2-1}{3}\right)\right).
\end{equation}
For the second term, 
\begin{equation}
    \left\vert \sum_{n = N+1}^\infty c_n \right\vert \leq \sum_{n = N+1}^\infty \left\vert  c_n \right\vert \leq \frac{2}{\pi}\int_{N}^\infty \frac{1}{n^2 - 1/4} \leq 
    \frac{2}{\pi}\int_{N}^\infty \frac{1}{(n - 1/4)^2} = \frac{2}{\pi} \frac{1}{N - 1/4}. 
\end{equation}
We see that the first term dominates over the second term. Now, we only need to re-arrange terms from \Cref{eq.52} to show
\begin{equation}
    \sup_{x \in [0, 1]} \lvert P_N(x) - e^{iwx} \rvert \in \Tilde{\mathcal O}\left( N^{-1}\right).
\end{equation}

We recover now the Frobenius norm from \Cref{le.one_gate}
\begin{equation}
        \norm{
        \R - e^{i w x\hat\sigma_z}
        }_F = \sqrt 2 \sqrt{\vert P_N(x) - e^{iwx}\vert^2 + \vert Q_N(x)\vert^2}.
    \end{equation}
By virtue of Fejér's theorem, the absolute value can be made arbitrarily small. Hence, in particular $\vert P_N(x) - e^{iwx}\vert^2 \leq \vert P_N(x) - e^{iwx}\vert \leq 1$.
It is immediate to see that
\begin{equation}
    \norm{
        \R - e^{i w x\hat\sigma_z}
        }_F \leq \sqrt{6 \vert P_N(x) - e^{iwx}\vert},  
\end{equation}
and subsequently
\begin{equation}
    \sup_{x \in [0, 1]} \norm{
        \R - e^{i w x\hat\sigma_z}
        }_F\in \Tilde{\mathcal O}\left( N^{-1/2}\right).
\end{equation}

Therefore, we can approximate the gate $e^{i w x\hat\sigma_z}$ with two steps. First, we apply $e^{i \pi x\hat\sigma_z}$ a number of times $N^\prime$ to reach $\vert \omega - N^\prime \pi \vert \leq \pi/2$. Second, we use Fejér's theorem to approximate $e^{i w x\hat\sigma_z}$, for $\vert w \vert \leq \pi/2$, with the errors specified in this proof. Hence, in order to approximate $e^{i w x\hat\sigma_z}$, to $\epsilon$ precision in the Frobenius norm, we need to apply 
\begin{equation}
    N \in \Tilde\Omega(w + \epsilon^{-2}) 
\end{equation}
gates. 
\end{proof}

For the next step, we subsequently apply \Cref{le.one_gate} to show that there exist parameters $\bm\theta^\prime, \bm\phi^\prime$ such that, for given $W, \bm\theta, \bm\phi$
\begin{equation}
    \uxfprime \approx \left(\prod_{j = 1}^L e^{i \theta^\prime_j\hat\sigma_z} e^{i \phi^\prime_{j}\hat\sigma_y}e^{i \sum_{k = 1}^m w_{j, k} x_k\hat\sigma_z} \right)e^{i \theta_{0}'\hat\sigma_z} e^{i \phi_{0}'\hat\sigma_y} e^{i \lambda' } = \uw.
\end{equation}
In a nutshell, we take our previous result to approximate a gate of the form $e^{i w_k x_k\hat\sigma_z}$. Then, we can subsequently apply these gates to approximately obtain $\prod_{k}e^{i x_k w_k\hat\sigma_z} = \exp{\left(i \sum_k x_k w_k\hat\sigma_z \right)}$, which are tunable weights. This step bridges the gap with models relying on tunable weights. We refer the reader to~\Cref{fig.qsp} for an illustration of this step.

\begin{figure}[b!]
\resizebox{\linewidth}{!}{
\begin{quantikz}[column sep=1mm]
    & \qw & \gate[][5cm]{\bm w \cdot \bm x} & \qw &  \push{~\approx~} &
     & \gate[style={fill = red!20!white}]{\theta_{m,N}}& \gate[style={fill = red!40!white}]{x_m} & \push{~\cdots~} & \gate[style={fill = red!40!white}]{x_m} & \gate[style={fill = red!20!white}]{\theta_{m,1}} & \push{~\cdots~}
     & \gate[style={fill = green!20!white}]{\theta_{2,N}}& \gate[style={fill = green!40!white}]{x_2} & \push{~\cdots~} & \gate[style={fill = green!40!white}]{x_2} & \gate[style={fill = green!20!white}]{\theta_{2,1}} & 
    & \gate[style={fill = blue!20!white}]{\theta_{1,N}}& \gate[style={fill = blue!40!white}]{x_1} & \push{~\cdots~} & \gate[style={fill = blue!40!white}]{x_1} & \gate[style={fill = blue!20!white}]{\theta_{1,2}} & \gate[style={fill = blue!40!white}]{x_1} & \gate[style={fill = blue!20!white}]{\theta_{1,1}} & 
\end{quantikz}}

\caption{Quantum circuit allowing the universality of quantum re-uploading models, with fixed encoding gates (in this figure represented by $\{x_1, x_2, \ldots, x_m$). By~\Cref{le.one_gate}, we show that gates of the form $e^{i w x\hat\sigma_z}$ can be approximated to arbitrary accuracy by a re-uploading circuit with fixed encoding gates. The successive application of these approximations allow for an approximation of $e^{i \bm w \cdot  \bm x\hat\sigma_z}$, where $\bm x$ is now multidimensional.}
\label{fig.qsp}
\end{figure}

\begin{lemma}\label{le.distances}
    Consider the re-uploading models $\uw$ from \Cref{eq.dru_tunable} and $\uxfprime$ from~\Cref{eq.dru_multi_fixed}. For any $\epsilon > 0$, $L$, $(\bm\theta,\bm\phi, W)$ there exists value $L^\prime \geq L$, $(\bm\theta^\prime, \bm\phi^\prime)$ satisfying
    \begin{equation}\label{eq.distances}
         \sup_{\bm x \in [0, 1]^m}\left\vert\Tr\left(\left(\uw - \uxfprime \right)\ket 0 \bra 0\right)\right\vert \leq \epsilon  
    \end{equation}
\end{lemma}

\begin{proof}
    We aim to bound the difference between the output functions of two unitaries built with and without tunable weights, explicitly $\uw$ and $\uxfprime$. First, Hölder's inequality allows us to write
    \begin{multline}
    \min_{\bm\theta, \bm\phi} \sup_{\bm x} \left\vert \Tr\left( \left(\uw - \uxfprime\right) \ket 0 \bra 0\right) \right\vert  \leq \\ 
    \min_{\bm\theta, \bm\phi} \sup_{\bm x} \norm{\uw - \uxfprime}_\infty,
    \end{multline}
    where the matrix norm is the $\infty$-Schatten norm. Both $\uw$ and $\uxfprime$ are constructed through layers. Following the rational of~\Cref{eq.R_gate}, we group the gates in $\uxfprime$ to approximate $e^{iwf\hat\sigma_z}$ in $\uw$. Then this is done, we can match the parameters in $(\bm\theta^\prime, \bm\theta), (\bm\phi^\prime, \bm\phi)$ to apply the same operations between encoding gates as in $\uw$. This fixes the parameters $\bm\theta^\prime, \bm\phi^\prime$, thus giving an upper bound to the previous equation. 

    We can now use the triangle inequality. Consider two unitaries given by $u U, v V$, where $u, v, U$ and $V$ are unitaries as well. Then
    \begin{equation}\label{eq.telescopit}
        \norm{u U - v V} = \norm{u U - v U + v U - v V} \leq \norm{u - v} + \norm{U - V}. 
    \end{equation}
    This procedure is repeated telescopically to the matrix 
    $\uw - \uxfprime$. We tune the parameters in $\uxfprime$ to exactly match those of $\uw$ in the gates between encoding gates, thus not contributing to the difference. For the steps involving $e^{i w_{j, k} x_k\hat\sigma_z}$, we approximate it with gates of the form $\R$. Thus, 
    \begin{equation}
        \norm{\uw - \uxfprime}_\infty \leq \sum_{j = 1}^{L^\prime} \sum_{k = 1}^m \sup_{x_k \in [0, 1]} \norm{\R - e^{i w_{j, k} x_k\hat\sigma_z}}_\infty.
    \end{equation}
    By virtue of \Cref{le.one_gate}, the Frobenius norm  $\Vert \R - e^{i w_{j, k} x_k\hat\sigma_z}\Vert_F$ can be made arbitrarily small. The Frobenius norm upper bounds the $\infty$-norm, thus each of the components in the sum can be made arbitrarily small. 
    
\end{proof}

For the final step, we just need to consider~\Cref{le.distances} and the triangle inequality. A model with tunable weights can approximate any function with arbitrary accuracy, and a fixed-weight model can approximate a model with tunable weights as well. Then, single-qubit fixed-weight models are universal approximators, at the expense of an overhead in depth.

\begin{theorem}
    Consider the model $\uxfprime$ from~\Cref{eq.dru_multi_fixed}. The sequence of output function families $\mathcal H = \{h_{L^\prime}^{\rm f} \}$, with  
    \begin{equation}
        h^{\rm f}_{L^\prime}(\bm x) = \bra 0 \uxfprime \ket 0
    \end{equation}
    is  universal with respect to multivariate continuous functions $f : [0, 1]^m \rightarrow \mathbb C$ with the constraint $\vert f(\bm x)\vert^2 \leq 1$, in the $\norm{\cdot}_\infty$ norm. 
\end{theorem}

\begin{proof}
    This corollary is an immediate consequence of~\Cref{le.distances}. The considered model can approximate a multivariate function $h_L(\bm x) = \bra 0 \uw \ket 0$ output by a model with tunable weights. The triangle inequality implies
    \begin{equation}
        \norm{h_{L^\prime}^{\rm f}(\bm x) - f(\bm x)} \leq \norm{h_L(\bm x) - h_{L^\prime}^{\rm f}(\bm x)} + \norm{h_{L}(\bm x) - f(\bm x)}.
    \end{equation}
    Each term can be made arbitrarily small. The first one by virtue of \Cref{le.one_gate}, and the second one following \Cref{th.dru_tunable}. Hence, universality is guaranteed. 
\end{proof}
This last theorem is analogous to the one written in the main text.

\subsection{Proof of~\Cref{cor.analysis}}\label{app.analysis}
\begin{proof}
To prove \Cref{cor.analysis}, we follow the same procedure as for the quantum re-uploading circuits. We will approximate $e^{i w x}$ with its Cesàro mean and merge all the functions together. The distance between a function $g_N(\bm x)$ from~\Cref{th.uat} and its discrete-weights approximation is given by
    \begin{equation}
        \sup_{x \in [0, 1]^m}\left\vert \sum_{j = 1}^L \gamma_j \prod_{k = 1}^m e^{i w_{j, k} x_k} - \sum_{j = 1}^L \gamma_j \prod_{k = 1}^m P_{N, \omega_{j, k}}(x_k) \right\vert \leq \\
        \sum_{j = 1}^L \vert \gamma_j\vert \sup_{x \in [0, 1]^m} \left\vert{\prod_{k = 1}^m e^{i w_{j, k} x_k} - \prod_{k = 1}^m P_{N, \omega_{j, k}}(x_k)}\right\vert, 
    \end{equation}
    with $P_{N, \omega_{j, k}}$ being the Cesàro means.  
    Since 
    \begin{equation}
        \sup_{x \in [0, 1]^m}\left\vert f(\bm x)\right\vert= \sup_{x \in [0, 1]^m}\left\vert e^{i \bm \alpha \cdot \bm x}f(\bm x)\right\vert, 
    \end{equation}
    we can perform an equivalent trick to the one in~\Cref{eq.telescopit} and find
    \begin{equation}
        \sup_{x \in [0, 1]^m}\left\vert\sum_{j = 1}^L \gamma_j \prod_{k = 1}^m e^{i w_{j, k} x_k} - \sum_{j = 1}^L \gamma_j \prod_{k = 1}^m \omega_{j, k}(x_k)\right\vert \leq 
        \sum_{j = 1}^L \vert \gamma_j \vert \sum_{k = 1}^m \sup_{x_k \in [0, 1]}\left\vert e^{i w_{j, k} x_k} - P_{N, \omega_{j, k}}(x_k)\right\vert.
    \end{equation}
    Since $P_{N, \omega_{j, k}}(x_k)$ is the Cesàro mean of $e^{i w_{j, k} x_k}$, up to degree $N$, we can recall Fejér's theorem to approximate the desired function with arbitrary precision. Hence, this construction provides universal functions in the assumptions of Fejér's theorem. 
\end{proof}

\subsection{Euler angles for $\mathcal{SU}(N)$}\label{app.euler}
We follow the construction in~\cite{tilma2002generalized} for Euler rotations of arbitrary dimensions. We define first the generators of the corresponding algebra $\mathfrak{su}(N)$ as generalized Gell-Mann matrices
\begin{align}
\hat g_3 & = \begin{pmatrix}
    1 & 0 & 0 &\cdots & 0 \\
    0 & -1 & 0 &\cdots & 0 \\
    0 & 0 & 0 & \cdots & 0 \\
      & \vdots &  & \ddots & \vdots \\
    0 & 0 & 0 & \cdots & 0
\end{pmatrix}    \\
\hat g_{(k-1)^2 + 1} & = \begin{pmatrix}
    \underbracket[1pt]{\begin{array}{ccc} 0 & \cdots & -i \\ 
    \vdots & \ddots & \vdots \\ 
    i & \cdots & 0\end{array}}_k & \begin{array}{cc}
         \cdots & 0 \\  & \vdots \\ \cdots & 0\end{array} \\
    \begin{array}{ccc} \vdots &  & \vdots \end{array}& \begin{array}{cc}\ddots & \vdots\end{array} \\
    \begin{array}{ccc} 0 & \cdots & 0 \end{array}& \begin{array}{cc}\cdots & 0 \end{array}
\end{pmatrix}  \\  
\hat g_{N^2-1} = \sqrt{\frac{2}{N^2 - N}} & \begin{pmatrix}
    1 & 0 & 0 &\cdots & 0 & 0 \\
    0 & 1 & 0 &\cdots & 0 & 0 \\
    0 & 0 & 1 & \cdots & 0 & 0 \\
      & \vdots &  & \ddots & & \vdots \\
    0 & 0 & 0 & \cdots & 1 & 0 \\
    0 & 0 & 0 & \cdots & 0 & -1
\end{pmatrix} 
\end{align}
There exists an Euler decomposition of arbitrary dimension by employing a recursive construction in which matrices in $\mathcal{SU}(N)$ are defined as matrices in $\mathcal{SU}(N-1)$ plus extra parameters. 
\begin{equation}
    U = \prod_{2 \leq k \leq N} e^{i \hat g_3 \theta_{2k - 3}} e^{i \hat g_{(k^2 - 1) + 1} \theta_{2k - 2}} [SU(N-1)] e^{i \hat g_{N^2 - 1} \theta_{N^2 - 1}} 
\end{equation}
Notice that this construction allocates all rotations constructed as exponentiations of diagonal matrices at the rightest part of the operations. Hence, we can justify our choice $V_m(\bm x)$ from~\Cref{eq.encoding_multidimension}. Such definition allows for a one-to-one identification of values $\bm x$ and the corresponding Euler angles for the diagonal elements. 

\subsection{Proof of~\Cref{th.multidimension}}\label{app.multidimension}
\begin{proof}

We begin by considering the $\bm x$-encoding gate
\begin{equation}
    V_m(\bm x) = \begin{pmatrix}
        e^{i \pi x_1/2} & 0 & \cdots & 0 & \cdots \\ 
        0 & e^{-i \pi x_1/2} & \cdots & 0 & \cdots \\ 
        0 & 0 & e^{i \pi x_2/2} & 0 & \vdots \\
        0 & 0 & 0 & e^{-i \pi x_2/2} & \vdots \\
        \vdots & \vdots & \vdots & \vdots & \ddots
    \end{pmatrix}
\end{equation}
We design now a permutation $\Pi_k$ that exchanges the positions $2j$ and $2j + 1$, except for $j = k$. An example of this permutation is given by 
\begin{equation}
    \Pi_1 V_m(\bm x) = \begin{pmatrix}
        e^{i\pi  x_1/2} & 0 & \cdots & 0 & \cdots \\ 
        0 & e^{-i \pi x_1/2} & \cdots & 0 & \cdots \\ 
        0 & 0 & e^{-i \pi x_2/2} & 0 & \vdots \\
        0 & 0 & 0 & e^{i \pi x_2/2} & \vdots \\
        \vdots & \vdots & \vdots & \vdots & \ddots
    \end{pmatrix}.
\end{equation}
These permutations are implementable with gates $R(\bm \theta, \bm \phi)$, since $R$ specifies any unitary gate, except for the relative phases among columns~\cite{tilma2002generalized}, and no phases are needed for the permutations. It is immediate to see
\begin{equation}
    R_{1}(\bm x) = V_m(\bm x) \Pi V_m(\bm x) = \begin{pmatrix}
        e^{i \pi x_1} & 0 & \cdots & 0 & \cdots \\ 
        0 & e^{-i \pi x_1} & \cdots & 0 & \cdots \\ 
        0 & 0 & 1 & 0 & \vdots \\
        0 & 0 & 0 & 1 & \vdots \\
        \vdots & \vdots & \vdots & \vdots & \ddots
    \end{pmatrix} = \begin{pmatrix}
    \begin{bmatrix}
        e^{i x_1} & 0 \\ 0 & e^{-i x_1}
    \end{bmatrix} & 0 \\ 
    0 & I
\end{pmatrix} = \begin{pmatrix}
    e^{i \pi x_1\hat\sigma_z} & 0 \\ 
    0 & I
    \end{pmatrix}.
\end{equation}
The construction of $R_1(\bm x)$ requires an overhead of $\mathcal O(1)$ encoding gates of the form $V_m(\bm x)$ to achieve it. The same condition holds for any other $R_k(\bm x)$. These blocks, up to permutations that can be reabsorbed in the parameterized gates, allow us to conduct the algorithms in~\Cref{th.dru_fixed} on the $2 \times 2$ upper-left corner. This can be extended to all coordinates. Notice that the relative phases cancel each other due to the matrix being applied the same number of times. Since universality is guaranteed in~\Cref{th.dru_fixed}, this construction allows for universality as well in the multi-qubit case. 
\end{proof}

\subsection{An extension to \Cref{th.multidimension}}\label{app.extension}
The proof above allows us to formulate an alternative theorem considering a different encoding gate, defined as
\begin{equation}\label{eq.encoding_multidimension}
    V^\prime_m(\bm x) = \sum_{k = 1}^m e^{i \pi x_k/m}\ketbra{k}{k}+ e^{-i \sum_{i = 1}^m \pi x_k/m}\ketbra{m + 1}{m + 1} = 
    \begin{pmatrix}
        e^{i \pi x_1/m} & 0 & 0 & \cdots & 0\\ 
        0 & e^{i \pi x_2/m} & 0 & \cdots & 0 \\ 
        0 & 0 & e^{i \pi x_3/m} & \cdots & 0 \\
        \vdots & \vdots & \vdots & \ddots & \vdots \\
        0 & 0 & 0 & \cdots & e^{-i \frac{\pi }{m}\sum_{k = 1}^m x_k}
    \end{pmatrix}
\end{equation}
We define the auxiliary permutations $\Pi_{k, j}$, for $j < k, k \leq m$, being circular permutations on all elements except for $k$, of distance $j$. Again, these permutations are implementable with gates $R(\bm \theta, \bm \phi)$, since $R$ specifies any unitary gate, except for the relative phases among columns~\cite{tilma2002generalized}, and no phases are needed for the permutations. An example of this permutation is
\begin{equation}
    \Pi_{1, 1}V^\prime_m(\bm x) = \begin{pmatrix}
        e^{i \pi x_1 / m} & 0 & 0 & \cdots & 0 \\ 
        0 & e^{-i \frac{\pi }{m} \sum_{k = 1}^m x_k} & 0 & \cdots & 0 \\
        0 & 0 & e^{i \pi x_2 / m} & \cdots & 0 \\
        \vdots   &  \vdots  & \vdots & \ddots & \vdots \\
        0 & 0 & 0 & \cdots & e^{i \pi x_{m - 1}/m} 
    \end{pmatrix}.
\end{equation}
We apply subsequent permutations until finding
\begin{equation}
    V^{\prime}_{\Pi_1}(\bm x) = \prod_{j = 1}^{m - 1} \Pi_{1, j}(V^{\prime}_m(\bm x)) = 
    \begin{pmatrix}
        e^{i x_1 \frac{m - 1}{m}} & 0 & 0 & \cdots & 0 \\ 
        0 & e^{-i x_1 / m} & 0 & \cdots & 0 \\
        0 & 0 & e^{-i x_1 / m} & \cdots & 0 \\
        \vdots   &  \vdots  & \vdots & \ddots & \vdots \\
        0 & 0 & 0 & \cdots & e^{-i x_1 / m} 
    \end{pmatrix} = 
    e^{- i x_1 / m}\begin{pmatrix}
        e^{i x_1} & 0 & 0 & \cdots & 0 \\ 
        0 & 1 & 0 & \cdots & 0 \\
        0 & 0 & 1 & \cdots & 0 \\
        \vdots   &  \vdots  & \vdots & \ddots & \vdots \\
        0 & 0 & 0 & \cdots & 1 
    \end{pmatrix}
\end{equation}
The next step is to apply the same permutation cycle over this permuted matrix to find
\begin{equation}
    V^{\prime}_{\Pi_1}(-\bm x) = \prod_{j = 1}^m \Pi_{2, j}(V^{\prime}_{\Pi_1}(\bm x)) = 
    e^{i \pi x_1 / m}\begin{pmatrix}
        1 & 0 & 0 & \cdots & 0 \\ 
        0 & e^{-i \pi x_1} & 0 & \cdots & 0 \\
        0 & 0 & 1 & \cdots & 0 \\
        \vdots   &  \vdots  & \vdots & \ddots & \vdots \\
        0 & 0 & 0 & \cdots & 1 
    \end{pmatrix}.
\end{equation}
We can identify 
\begin{equation}
R_{1}(\bm x) = V^{\prime}_{\Pi_1}(-\bm x) V^{\prime}_{\Pi_1}(\bm x) = \begin{pmatrix}
    \begin{bmatrix}
        e^{i \pi x_1} & 0 \\ 0 & e^{-i \pi x_1}
    \end{bmatrix} & 0 \\ 
    0 & I
\end{pmatrix} = \begin{pmatrix}
    e^{i \pi x_1\hat\sigma_z} & 0 \\ 
    0 & I
\end{pmatrix}.
\end{equation}
The construction of $R_1(\bm x)$, or equivalent $R_k(\bm x)$ requires an overhead of $\mathcal O(m^2)$ encoding gates $V^{\prime}_m(\bm x)$. 

From this point, we can repeat the steps in \Cref{app.multidimension} and formulate the following theorem. 

\begin{theorem}\label{th.multidimension2}
Let $\ufm$ be the gate defined in~\Cref{def.dru_fixed_large}, with encoding gate $V^\prime_m(\bm x)$. The family of output functions $\mathcal H = \{h_{L} \}$, with  
    \begin{equation}
        h_L(\bm x) = \bra 0 \ufm \ket 0
    \end{equation}
    is universal with respect to the norm $\norm{\cdot}_\infty$ for all continuous complex functions $f(\bm x), \norm{f(\bm x)}_\infty \leq 1$, for $\bm x \in [0, 1]^m$. This model has overhead $\mathcal O(m^2)$ in depth as compared to the single-qubit case. 
\end{theorem}

The main difference of this theorem with respect to \Cref{th.multidimension} is that this case requires one qubit less. However, the overhead in depth as compared to the single-qubit case is beneficial to \Cref{th.multidimension}. 

\end{document}